\documentstyle[preprint,aps,tighten]{revtex}

\def\be{\begin{eqnarray}}
\def\ee{\end{eqnarray}}

\def\L{\langle}
\def\R{\rangle}

\begin{document}
\title{Spin-polarized states of nuclear matter}
\author{
W. Zuo$^{1,2}$, 
U. Lombardo$^{2,3}$ and C. W. Shen$^{2}$}
\address{
$^{1}$ Institute of Modern Physics, Lanzhou, China\\
$^{2}$INFN-LNS, 44 Via S.~Sofia, I-95123 Catania, Italy \\
$^{3}$ Dipartimento di Fisica, 57 Corso Italia, I-95129 Catania, Italy }
\begin{titlepage}
\maketitle
\setcounter{page}{1}
\vskip 4.0 truecm
\centerline{\bf Abstract}

The equations of state of spin-polarized nuclear matter and pure neutron matter 
are studied in the framework of the Brueckner-Hartree-Fock theory including 
a three-body force. 
The energy per nucleon $E_A(\delta)$ calculated in the full range of spin 
polarization ${\delta} = \frac{\rho_{\uparrow}-\rho_{\downarrow}}{\rho}$ 
for symmetric
nuclear matter and pure neutron matter fulfills a parabolic law. In both cases
the spin-symmetry energy is calculated as a function of the baryonic density
along with the related quantities such as  the
magnetic susceptibility and the Landau parameter $G_0$. 
The main effect of the three-body force is to strongly
reduce the degenerate Fermi gas magnetic susceptibility even more than the 
value with only two body force.
The EOS is monotonically increasing with the density for all spin-aligned
configurations studied here so that no any signature is found for a
spontaneous transition to a ferromagnetic state.

%
%
%
%

\end{titlepage}

\section{Introduction}

Studies of spin-polarized nuclear and neutron matter have been mainly 
focussed on 
the possible onset of a ferromagnetic transition in the neutron star 
core.
This transition could explain in fact the high intensity magnetic 
fields ($10^{12}$ gauss) estimated
from the timing observations in pulsars and magnetars~(for a review see 
Ref.~\cite{REIS}). 
Besides this exciting issue the motivation for such studies can be based on
a more general context of nuclear physics. 

First of all, from the change of the energy per nucleon with the spin 
polarization one may extract a theoretical prediction for the spin symmetry 
energy, whose empirical value is so far quite uncertain. The nuclear matter 
instability against spin fluctuations is driven by the Landau parameter $G_0$ 
which is determined from the spin-symmetry energy. 
The value of this parameter is 
still a largely controversial topic and no agreement exists among the 
different approaches to the theory of nuclear matter~\cite{LOMB}. 
Experimental information could come from spin giant resonances, which have not 
yet been clearly observed. Information could also come from heavy-ion 
collisions as soon as polarized heavy targets become available.     

The second issue related to the study of spin-aligned states of nuclear matter
is the propagation of neutrinos in neutron stars. It has been shown that the
neutrino mean free path is strongly affected by the magnetic susceptibility.
The latter is sizeably suppressed by the strong correlations in nuclear matter
and, as a consequence, the mean free path might change  sizeably and, 
eventually drop to zero in the presence of a ferromagnetic 
transition~\cite{MARG1,MARG2}.
 
There is a guess that the ferromagnetic transition could be a relativistic
effect due to $\pi$-exchange 
and in fact all calculations, based on the relativistic mean field approch,
  predict this transition to occur in dense matter~\cite{BERN,MARU}. 
On the other hand, 
non-relativistic approaches~\cite{HAEN,CUGN,VIDA,BOMB} 
do not support such a transition except  
Hartree-Fock calculations with phenomenological 
Skyrme-like forces~(for a review see Ref.~\cite{MARG2}). 
This aspect cannot be disconnected from the problem of the in-medium 
nucleon-nucleon (NN) 
force, which is poorly known in dense matter due to the 
lacking of empirical constraints far above the saturation density. 
However important relativistic effects can be incorporated into the 
effective interaction via the three-body force associated with a virtual 
nucleon-antinucleon excitation~\cite{LEJE}. 
Moreover non-relativistic calculations including only two-body forces 
miss the empirical saturation point of nuclear matter~\cite{BAL}. 
So it seems worthwhile to investigate the spin-aligned states 
of nuclear and neutron matter in the non-relativistic Brueckner 
theory with three-body forces. They contain not only the 
above mentioned relativistic contributions but also nucleonic excitations 
which decisively enhance the agreement between theoretical 
and empirical saturation density~\cite{ZUOW}. 

\section{Formalism}

The spin and isospin asymmetric nuclear matter (ANM) 
consists of spin-up neutrons ($n\uparrow$), spin-down neutrons ($n\downarrow$), 
spin-up protons ($p\uparrow$) and spin-down protons ($p\downarrow$) in 
different density states: $\rho_{n\uparrow}$, $\rho_{n\downarrow}$,
$\rho_{p\uparrow}$, and $\rho_{p\downarrow}$, respectively. Therefore four 
parameters are required to specify a given configuration of spin and 
isospin ANM. The Fermi momenta of the 
four components are generally different from each other, and related 
to their respective densities by the following relation: 
$$
\rho_{\lambda}=\frac{1}{6\pi^2}(k^{\lambda}_F)^3
$$
where $\lambda$ denotes the $z$-components of isospin and spin, i.e., 
$\lambda=(\tau_z,\sigma_z)$. 
Instead of $\rho_{\lambda}$, one can use the 
following four parameters to identify a given spin and isospin state, 
$$
\beta=\frac{\rho_n-\rho_p}{\rho}, \ \ \ 
\delta_n=\frac{\rho_{n\uparrow}-\rho_{n\downarrow}}{\rho_n}, \ \ \ 
\delta_p=\frac{\rho_{p\uparrow}-\rho_{p\downarrow}}{\rho_p}, \ \ \ 
$$
where $\rho$, $\rho_n$, and $\rho_p$ are total density, neutron density and
proton density, respectively. The ratio $\beta$ is the isospin asymmetry 
parameter and $\delta_n$ and $\delta_p$ are the spin asymmetry parameters 
for neutrons and protons, respectively. 

The starting point of the Brueckner-Bethe-Goldstone (BBG) approach is the 
reaction $G$-matrix. The $G$-matrix incoporates strong short-range 
correlations in nuclear medium by means of the infinite ladder diagram 
summation of the bare NN interaction. It 
satisfies the Bethe-Goldstone equation. The latter can be expressed 
for the spin-isospin ANM in the total angular-momentum basis as follows,  
$$
\begin{array}{lll}
G^{TSJ,\lambda\lambda'}_{LL'}(\omega , P; q, q'; \rho, 
\beta, \delta_n, \delta_p) 
& = & \displaystyle 
v^{TSJ}_{LL'}(q, q') + \frac{2}{\pi}\sum_{L''}
\int q''^2 {\rm d} q'' 
v^{TSJ}_{LL''}(q, q'') \\
\nonumber\\
&\times & \displaystyle 
\frac{\L Q^{\lambda\lambda'}(q'',P)\R }
{\omega - e^{\lambda\lambda'}_{12} (q'',P) + i\eta}
G^{TSJ,\lambda,\lambda'}_{L''L'}(\omega , P; q'', q'; 
\rho, \beta, \delta_n, \delta_p) 
\end{array}
$$
where $v^{TSJ}_{LL'}(q, q')$ are the partial 
wave components of the NN interaction, 
$\vec{P}=\vec{k}_1+\vec{k}_2=\vec{k}_1'+\vec{k}_2'$ is the total momentum, 
$\vec{q}=\displaystyle 
\frac{\vec{k}_1-\vec{k}_2}{2}$ and 
$\vec{q}\ '=\displaystyle 
\frac{\vec{k}_1'-\vec{k}_2'}{2}$ the relative momenta of the two particles 
in their initial state and final state, respectively. 
The Pauli operator $ Q^{\lambda\lambda'}(\vec{q}\ '', \vec{P} ) $ 
and the energy denominator 
$ e_{12}^{\lambda \lambda'} (\vec{q}\ '', \vec{P} ) = 
\epsilon^{\lambda} (k''_1) + \epsilon^{\lambda'}(k''_2)$ 
have been angle averaged in order to remove the coupling 
between different channels $\alpha = \{JST\}$. It is worth noticing 
that the different components of the $G$-matrix differ in general 
from each other due to the dependence of the Pauli operator and energy 
denominator on the spin-isospin configuration 
($\lambda, \lambda\ '$). The single particle 
energy is given by 
$\epsilon^{\lambda} (k)= \hbar^2 k^2/2m + U^{\lambda}(k)$. 
The continuous choice for the auxilary potential $U^{\lambda}(k)$ 
is adopted in the present calculations since, on the one hand, it 
has been shown to provide a much faster convergence of the 
hole-line expansion than the gap choice~\cite{SONG}, on the other hand, it 
decribes physically the single-particle potential felt by a nucleon 
in nuclear medium. In the continuous choice, 
$U^{\lambda}(k)$ is the real part of the on-shell mass 
operator, i.e., 
$$
U^{\lambda}(k) = Re \sum_{\vec{k}', \lambda' } 
n^{\lambda'}(k') \L \vec{k}\lambda,\vec{k}\ '\lambda' 
| G [\omega=\epsilon^{\lambda} (k) + \epsilon^{\lambda'} (k'), P] | 
\vec{k}\lambda,\vec{k'}\lambda' \R_A,
$$
where the subscript $A$ denotes antisymmetrization. 
For spin and isospin ANM, it is convenient to split it into 
two contributions as 
$$
U^{\lambda} = U^{\lambda\lambda} + U^{\lambda\lambda '},~~~(\lambda ' 
\ne \lambda).
$$
Each individul contribution is calculated by casting it into the partial wave 
expansion, 
$$
\begin{array}{lll}
U^{\sigma_z\tau_z,\sigma'_z\tau'_z}(k) 
& = & \displaystyle 
\int_0^{k_F^{\sigma_z'\tau_z'}}{\rm d}^3k' \sum_{TSJ}\sum_{LL'}\sum_{S_zT_z}
i^{L-L'}\left[C(\frac{1}{2}\sigma_z\frac{1}{2}\sigma_z'|SS_z)\right]^2
\left[C(\frac{1}{2}\tau_z\frac{1}{2}\tau_z'|TT_z)\right]^2 \\
& \times & \displaystyle 
\sum_{M_L} C(L'M_LSS_z|JM_L+S_z)C(LM_LSS_z|JM_L+S_z)
Y^*_{L'M_L}(\hat{q})Y_{LM_L}(\hat{q}) \\
& \times & \displaystyle 
2G^{TSJ,\lambda\lambda'}_{LL'}(\omega,P,q,q;\rho, \beta, 
\delta_n, \delta_p). 
\end{array}
$$
The summation over partial wave states is physically constrained by the 
selection rule $S+T+L={\it odd}$ due to the Pauli principle and 
consequently the antisymmetrization of the $G$-matrix simply implies 
multiplication by a factor of 2 for the allowed partial wave channels. 
For spin symmetric case ($\delta_n=\delta_p=0$), 
a spin-up neutron (proton) has the same Fermi momentum 
as a spin-down neutron (proton) and 
thus the single-particle potential felt 
by a nucleon does not depend on the direction of its spin. 
The  summation on the spins of the two particles in 
the final state and the average of that in the initial state remove 
the non-diagonal contributions in angular-momentum 
from the single-particle potential. One easily finds 
$$
\begin{array}{lll}
U^{\tau_z\tau_z'}(k) & = & \displaystyle 
\frac{1}{2} 
\sum_{\sigma_{z}\sigma'_{z} } 
U^{\sigma_z\tau_z,\sigma'_z\tau'_z}(k) \\
& = & \displaystyle 
\frac{1}{2} 
\int_0^{k_F^{\tau_z'}}{\rm d}^3k' \sum_{TSJL}\sum_{T_z}
\left[C(\frac{1}{2}\tau_z\frac{1}{2}\tau_z'|TT_z)\right]^2 
\frac{2J+1}{4\pi} 
2G^{TSJ,\tau_z\tau_z'}_{LL}(\omega,P,q,q;\rho, \beta). 
\end{array}
$$
For spin-asymmetric but isospin-symmetric nuclear matter, 
we have $\beta = 0$ and $\delta_n=\delta_p=\displaystyle 
\frac{\rho_{\uparrow} 
-\rho_{\downarrow}}{\rho}=\delta$. In this case, the 
single particle potential becomes, 
$$
\begin{array}{lll}
U^{\sigma_z,\sigma'_z} 
& = & \displaystyle 
\int_0^{k_F^{\sigma_z'}}{\rm d}^3k' \sum_{TSJ}\frac{2T+1}{2}
\sum_{LL'}\sum_{S_z} i^{L-L'}
\left[C(\frac{1}{2}\sigma_z\frac{1}{2}\sigma_z'|SS_z)\right]^2 
2G^{TSJ,\sigma_z\sigma_z'}_{LL'}(\omega,q,q;\rho, \delta) \\
& \times & \displaystyle 
\sum_{M_L} C(L'M_LSS_z|JM_L+S_z)C(LM_LSS_z|JM_L+S_z)
Y^*_{L'M_L}(\hat{q})Y_{LM_L}(\hat{q}). 
\end{array}
$$
The present calculations will mainly consider this spin-polarized nuclear 
matter as well as the spin polarized neutron matter.
 
\section{Results and conclusions}

We performed some calculations within the BHF self-consistent approach 
above described. The Argonne $V_{18}$ force is adopted as bare two-body 
interaction. This has been implemented by a microscopic three-body force
, which is described in detail 
in Ref.~\cite{MATH} together with the average procedure to transform 
it into an effective two body force.
In Fig.~1 the energy shift per nucleon $E_A(\delta,\rho)-E_A(0,\rho)$ 
in symmetric nuclear matter is reported as a function of the square of 
spin polarization $\delta^2$ for a set of densities. 
Due to the linear dependence on $\delta^2$, also 
reported in Refs.~\cite{VIDA,BOMB} one can write:
$$
 E_A (\delta,\rho) = E_A (0,\rho) +  {\cal E}_{sym}(\rho) \delta^2 ,
$$
i.e. the spin dependence of the energy per nucleon can be simply 
expressed in terms of spin-symmetry energy 
$$
{\cal E}_{sym}(\rho) = 
\frac{1}{2} \frac{\partial^2 E_A(\delta,\rho)}{\partial\delta^2}
$$
in the density range here considered.
                             
The $\delta^2$-law is mainly due to the BHF approximation (two-hole line terms
only). The same behavior is in fact exhibited by the energy vs. isospin either
\cite{ASY}. 
The effects of three-hole line terms are rather 
small when adopting the continuous choice for the auxiliary potential 
~\cite{SONG}; this choice is also adopted for the present calculations. 
The slope of the energy shift is monotonically increasing with 
density so that no signature for a ferromagnetic phase transition in symmetric 
nuclear matter is expected. The effect of three-body force 
is to enhance this slope 
for densities above the saturation point. This effect is more clearly 
shown in the plot of ${\cal E}_{sym}$ vs. density in Fig.~1. 

The magnetic susceptibility has been also calculated from the 
spin-symmetry energy
$$
  \chi = \frac{{\bar{\mu}}^2\rho} {2{\cal E}_{sym}},  
$$
where $\bar{\mu}$ is the average of neutron and proton magnetic moments
(in neutron matter $\bar\mu$ is the exact neutron magnetic moment).
Usually one calculates the ratio of $\chi$ to $\chi_F$,
$\chi_F$ being the magnetic susceptibility for a degenerate free 
Fermi gas.
The effect of strong correlations in nuclear matter due to the two-body 
force
is a reduction of $\chi$ with respect to $\chi_F$. This reduction 
increases with
density up to a factor of 0.3 at $\rho = 0.8 fm^{-3}$. 
The above result is common to most Bruckner 
calculations~\cite{CUGN,LEJE,VIDA,BOMB}. More pronounced 
is the queching due to three-body force. 

Fig.~2 also shows the Landau parameter $G_0$ describing the spin density 
fluctuations in the effective interaction. $G_0$ is simply related to the 
spin-symmetry energy or, equally, to the magnetic susceptibility by the 
relation
$$
            \frac{\chi}{\chi_F} = \frac{m^*}{1 + G_0}
$$
where $m^*$ is the effective mass. 
A magnetic instability would require $ G_0 < -1 $ which is analogous to the
condition $F_0 < -1$ for the mechanical instability giving rise to the 
liquid-vapour phase transition. But, the value of $G_0$ vs. density 
from the BHF calculations is 
always positive and monotonically increasing up to the highest density. 
The three-body force pushes up the curve of $\chi$. This result is 
in strong disagreement with
the prediction with Skyrme forces. This is not a complete surprise since
Skyrme forces are only well suited in the proximity of the empirical 
saturation point.
Astonishing is the strong disagreement on this respect with the relativistic
approaches because the three-body forces contain already important 
relativistic effects~\cite{MATH}. 
The accurate knowledge of $G_0$ should lead to reliable predictions on the
spin and spin-isospin giant modes as well as spin-spin part of the optical 
potential~\cite{CUGN}. 

The above calculations have been also repeated for the case of pure neutron
and reported in Fig.~2. 
The same conclusions can be 
drawn as to the absence of the ferromagnetic phase transition and the 
quenching of the magnetic susceptibility 
caused by the strong correlations from the two- and three-body forces. 
This quenching should have a 
strong influence on the neutrino propagation in dense matter such as 
supernovae and neutron stars. In the case of the transition to a ferromagnetic
state it has been shown that the mean free path could drop to zero \cite{MARG1}
that could have remarkable consequences as, for instance, on the neutron star 
cooling.

\newpage
\noindent

\begin{figure}[htb]
\begin{center}
\end{center}
\caption[]{Uper windows: The EOS of spin-asymmetric nuclear matter 
as a function of spin-asymmetry at five values of density, 
predicted by Brueckner Hartree-Fock calculations adopting 
pure AV$_{18}$ two-body force (right-uper window) and 
AV$_{18}$ plus the TBF(left-uper window). Right-Lower window: 
density dependence of spin-symmetry energy for both cases with 
the TBF (solid curve) and without the TBF(dash curve). 
Left-lower window: Magnetic suscepbility $\chi/\chi_0$ 
(curves with symbol) and 
Landau parameter $G_0$ (curves without symbol) as functions of density. 
}
\end{figure}
\begin{figure}[htb]
\begin{center}
\end{center}
\caption[]{The same as Fig.1 for spin-polarized neutron matter.}
\end{figure}

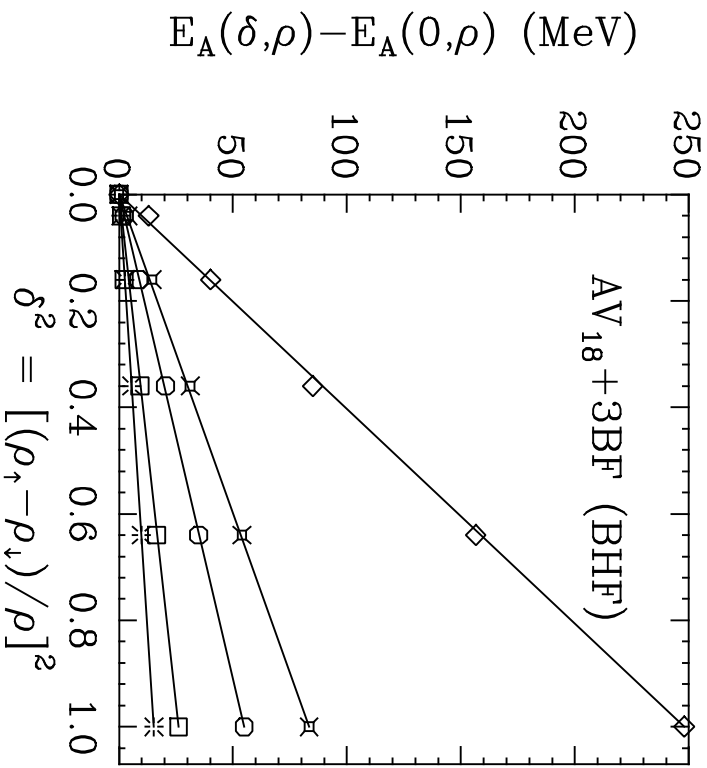
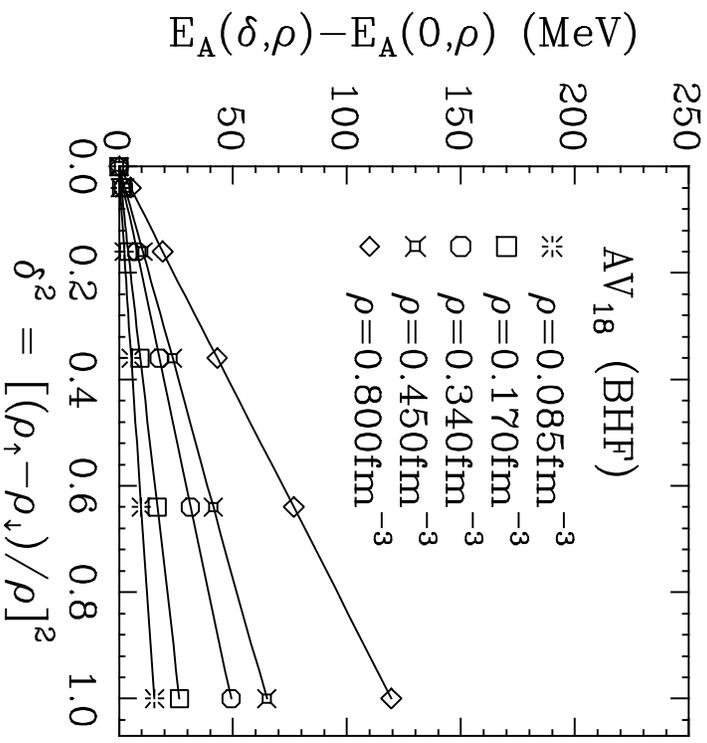
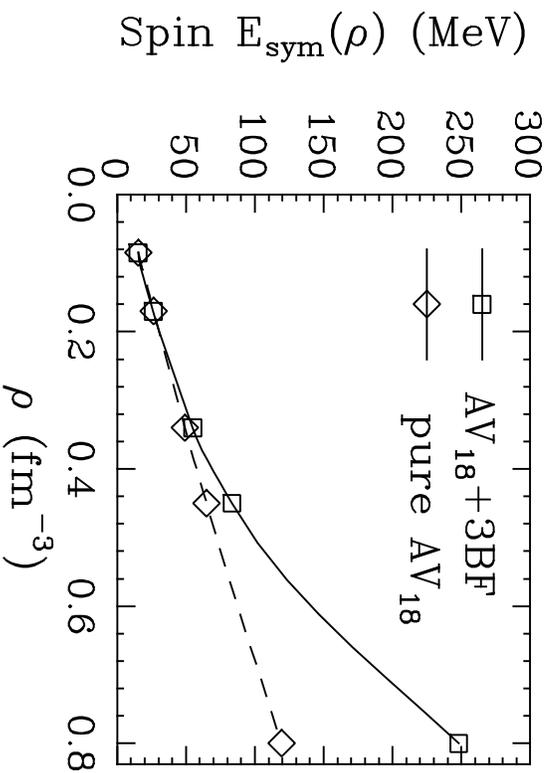
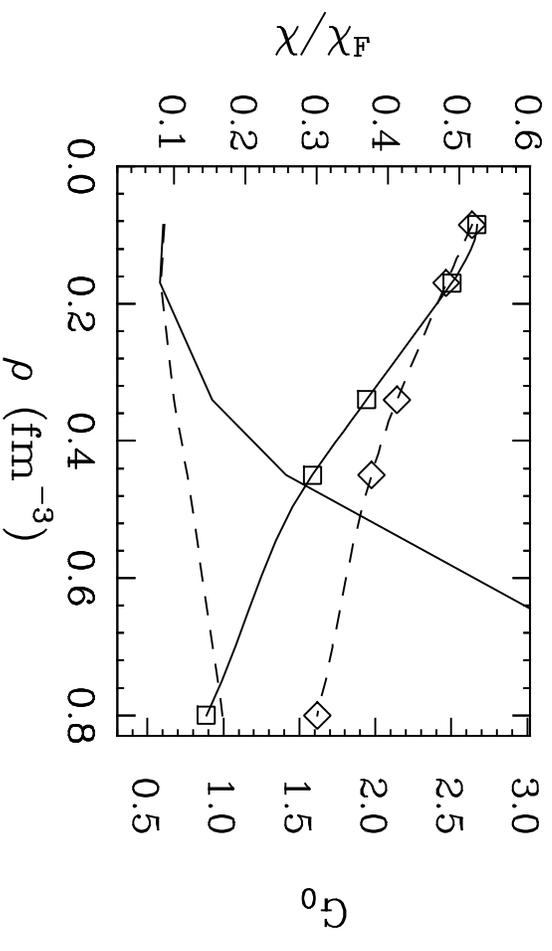
\end{document}